\documentclass[conference]{IEEEtran}
\IEEEoverridecommandlockouts
\usepackage{cite}
\usepackage{amsmath,amssymb,amsfonts}
\usepackage{graphicx}
\usepackage{textcomp}
\usepackage{xcolor}
\usepackage{hyperref}
\usepackage{algpseudocode,algorithm}
\def\BibTeX{{\rm B\kern-.05em{\sc i\kern-.025em b}\kern-.08em
    T\kern-.1667em$\lower.7ex\hbox{E}$\kern-.125emX}}
\begin{document}

\title{Detection of large exact subgraph isomorphisms with a topology-only graphlet index built using deterministic walks}

\author{
    \IEEEauthorblockN{
        Patrick Wang\IEEEauthorrefmark{2}\thanks{\IEEEauthorrefmark{2}These authors contributed equally.}
    }
    \IEEEauthorblockA{
        \textit{Department of Computer Science} \\
        \textit{University of California, Irvine}\\
        Irvine, United States \\
        wangph1@uci.edu
    }
    \and
    \IEEEauthorblockN{
        Henry Ye\IEEEauthorrefmark{2}
    }
    \IEEEauthorblockA{
        \textit{Department of Computer Science} \\
        \textit{University of California, Irvine}\\
        Irvine, United States \\
        hanweny@uci.edu
    }
    \and
    \IEEEauthorblockN{
        Wayne Hayes\IEEEauthorrefmark{1}\thanks{\IEEEauthorrefmark{1}Corresponding author.}
    }
    \IEEEauthorblockA{
        \textit{Department of Computer Science} \\
        \textit{University of California, Irvine}\\
        Irvine, United States \\
        whayes@uci.edu
    }
}

% TODO: co-first author thing
% To read: abstract, evaluation

\maketitle
\begin{abstract}
We introduce the first algorithm to perform topology-only local graph matching (a.k.a. local network alignment or subgraph isomorphism): BLANT, for Basic Local Alignment of Network Topology. BLANT first creates a limited, high-specificity index of a single graph containing connected $k$-node induced subgraphs called $k$-graphlets, for $k$=6--15. The index is constructed in a deterministic way such that, if significant common network topology exists between two networks, their indexes are likely to overlap. This is the key insight which allows BLANT to discover alignments using only topological information. To align two networks, BLANT queries their respective indexes to form large, high quality local alignments. BLANT is able to discover highly topologically similar alignments ($S^3 \ge 0.95$) of up to 150 node-pairs for which up to ~50\% of node pairs \textit{differ} from their ``assigned'' global counterpart. These results compare favorably against the baseline, a state-of-the-art local alignment algorithm which was adapted to be topology-only. Such alignments are 3x larger and differ ~30\% more (additive) more from the global alignment than alignments of similar topological similarity ($S^3 \ge 0.95$) discovered by the baseline. We hope that such regions of high local similarity and low global similarity may provide complementary insights to global alignment algorithms.
\end{abstract}


\section{Introduction}
\label{sec:intro}
% our solution is to do a hybrid between small seeds and aligning whole graphs at a time, using the unique capabilities of BLANT and our unique insights about indexing and techniques to allow topological similarity to predict functional similarity
Network alignment, or graph matching, is a common graph-mining problem that involves finding topologically similar regions between two or more networks. This problem has applications in a wide variety of fields \cite{emmert2016fifty}. These include bioinformatics \cite{clark2014comparison}, linguistics \cite{Dehmer2011}, neuroscience \cite{Sommerfeld1994}, social networks \cite{metaDiagramAlignment}, and image recognition \cite{HSIEH2008401}. Being a generalization of subgraph isomorphism, it is NP-hard \cite{GareyJohnson}, so many heuristic algorithms exist to approximately solve it. In this paper we consider only the case of aligning \textit{two} networks, although our algorithm can easily be extended to the multiple network case.

Network alignment can be either local or global \cite{meng2016local}. Global network alignment algorithms map all the nodes from the smallest network to the larger one(s), which is useful for quantifying the overall similarity between the networks. By contrast, local alignment algorithms find smaller conserved regions between networks with high local similarity but not necessarily high global similarity. The two approaches yield complementary insights \cite{meng2016local, GRAAL, GHOST}, motivating the need to research both types of algorithms.

In addition to the distinction between global and local, another key characteristic of a network alignment algorithm is whether it uses information beyond the graph's topology. For example, an alignment algorithm may rely heavily on domain-specific node attributes like usernames in a social network \cite{LIU2018318}, protein sequences in a biological network \cite{mina2012alignmcl}, or entity attributes in a knowledge graph \cite{RDFGraphAlignment}. Some algorithms do not use node attributes but do rely on pre-aligned seed nodes \cite{korula2014,kazemi2015}. However, these attributes or seeds are difficult to obtain and reduce the algorithms' generalizability (cf. \S\ref{sec:local_with_side}).

On the other end of the spectrum, there are alignment algorithms which rely solely on topological information \cite{MamanoHayesSANA,regal2018,dana2019}. These algorithms remove the significant cost of gathering domain-specific attributes or seeds and can also generalize to networks in any domain. Additionally, such algorithms may discover unique insights hidden in the topology of the graph itself \cite{wang2022sana,WeBeat}. However, the topology-only algorithms listed above are all global alignment algorithms. To the best of our knowledge, there do not currently exist any local alignment algorithms which use only topological information. BLANT seeks to fill this gap in the state of the art.

Designing a topology-only local alignment algorithm is non-trivial, because it is difficult to adapt existing techniques from either topology-only global aligners or non-topology-only local aligners. Topology-only global alignment algorithms operate on the entire graph as a whole, such as by randomly permuting the alignment in an evolutionary manner or by aligning and embedding of both graphs (see \S\ref{sec:topo_only_global} for more discussion). These techniques optimize for global similarity and thus miss small regions of high local similarity \cite{meng2016local}. The technique of creating a global alignment first and then mining it for local alignments, used by \cite{milano2018glalign}, does not solve this fundamental issue either, which is why we consider such algorithms to be global alignment algorithms. In summary, it is necessary to develop an approach which exclusively operates locally, not globally, in order to discover locally conserved regions which are not discovered by global alignment.

However, local approaches are intrinsically difficult to develop using only topological information because of a fundamental tension between complexity and information specificity. If only topology is used, individual nodes are indistinguishable; only larger structures contain enough information specificity in order to inform alignment. However, as the size of the desired structures increases, the number of said structures in the graph grows exponentially \cite{Santoso2020EfficientEO}, resulting in unacceptable complexity. Existing non-topology-only local aligners do not face this tension, making it difficult to adapt such algorithms to be topology-only. With node attributes, two individual nodes may already include enough information to be aligned. With seeds, the algorithm may ``collapse'' a significant amount of complexity by ``percolating'' the alignment to its neighbors.

In order to overcome this tension, BLANT uses the innovative approach of \textit{deterministically} mining graphlets—small induced subgraphs of a larger network—from a graph to create a graphlet index. BLANT's key insight is that, if we assume the networks \textit{have} similarity, a deterministic mining algorithm will likely extract a similar subset of graphlets across them. This idea resolves the tension between complexity and information, as we are able to mine a tractable set of graphlets that are large enough where their topology alone is usable as an identifier. Specifically, we mine graphlets of 6-15 nodes with BLANT, while the exhaustive enumeration algorithm ORCA \cite{orca} only outputs graphlets of 5 nodes as anything more is intractable. Additionally, the inclusion of a graphlet into the deterministically mined set is, in itself, important information. Even if a certain graphlet shape appears hundreds of times throughout two graphs, if that shape only appears once in both deterministic indexes, the two graphlets are likely to be counterparts.

This deterministic approach is in contrast to existing graph indexing techniques which either exhaustively enumerate structures (such as paths, trees, graphlets, etc.) or stochastically sample a large subset of structures in a graph \cite{katsarou2015}. The key difference between BLANT's index and existing indexes is that BLANT performs an existence check, while other indexes are used for non-existence checks (see \S\ref{sec:subgraph_querying} for additional discussion). Because the goal of other indexes is to rule \textit{out} the existence of a structure in a graph, they need to be fairly exhaustive. As a result, they are time and memory intensive: the fastest algorithm evaluated in \cite{katsarou2015} took 3 hours and 50GB to index a graph of 2000 nodes. On the other hand, because BLANT's index is decidedly non-exhaustive, we only require 1 hour and 25MB to index a graph that is 10x larger: 20000 nodes. BLANT uses non-exhaustion as a strength, because the fact that a given graphlet is included in an extremely narrow index at all is valuable information when performing alignment, as described in the previous paragraph.

After aligning a set of deterministically mined graphlets, BLANT then merges the aligned graphlets of 6-15 nodes together in order to create large graphlets. BLANT is able to discover highly topologically similar alignments ($S^3 \ge 0.95$) of up to 150 node-pairs for which up to ~50\% of node pairs \textit{differ} from their ``assigned'' global counterpart on a majority of the protein-protein interaction (PPI) network pairs in the Integrated Interactions Database (IID) \cite{kotlyar2015integrated}. Additionally, BLANT is able to output $S^3 \ge 0.95$ alignments of ~50 nodes for temporal network pairs from the Stanford Network Analysis Project database with 1\% - 5\% noise level \cite{snapdatabase} with ~50\% global dissimilarity. There are no direct competitors to BLANT as it is the first topology-only local alignment algorithm, but these results outperform those gathered by a version of a state-of-the-art local alignment algorithm (AlignMCL) which was adapted to be topology-only by \cite{meng2016local}. Compared to topology-only AlignMCL, BLANT's results are 3x larger and have ~30\% (additive) more global dissimilarity on the IID network pairs. They are ~1.5x larger and have ~30\% more global dissimilarity on the temporal network pairs. BLANT achieves this with a lower time complexity – $O(n)$ on a database of networks to AlignMCL's $O(n^2)$ with lower wall-clock time on our testbed (\S\ref{sec:runtime_and_storage}).

% existing approaches use global alignment as a precursor for local alignment through the alignment graph, but we think this has some issues. first, this lessens the benefit of local alignment because the results will be more similar to that of global alignment (less complementary). second, global alignment is not built for finding large clusters of similarity (and the MCL clustering algorithm is unable to do so as demonstrated). we take a completely different approach which also resolves this tension

% Many recent advances in network analysis have focused on using randomness, using techniques like network propagation ([https://link.springer.com/chapter/10.1007/978-3-030-57173-3\_9]), stochastic graph walking ([]), or Markov chains ([]). Even BLANT itself started off as a random sampling algorithm (which we now refer to as BLANT-sample). BLANT-seed, however, uses an innovative and counter-intuitive approach of creating a \textit{deterministic index} on two networks which it mines for seeds.

% RANDOM STUFF
% mention how sana can find alignments with no sequence similarity, which is actually a great use case for sana as it gives orthogonal alignments
% make sure to mention the benefits of determinism over randomness in the intro/background

\section{Background}
\subsection{Graphlets, Orbits, Ambiguity}
\label{sec:graphlets_orbits_ambiguity}
\begin{figure}
    \centering
    \includegraphics[width=0.47\textwidth]{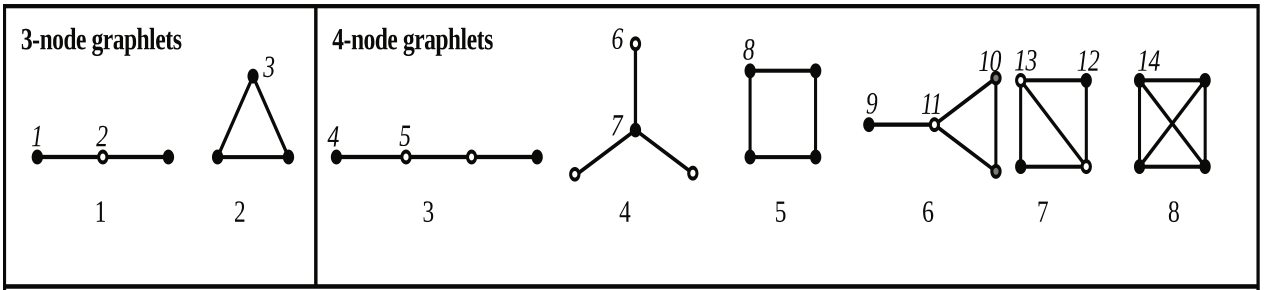}
    \caption{ All graphlets of sizes k = 3 and 4 nodes, and their automorphism orbits; within each graphlet, nodes of equal shading are in the same orbit. This figure is taken from \cite{Melckenbeeck1EtAl2016}. (Note that BLANT uses different, automatically-generated IDs than \cite{Melckenbeeck1EtAl2016}.) }
    \label{fig:graphlets}
\end{figure}

Graphlets are defined as ``small'' induced subgraphs of a network. They were originally exhaustively enumerated up to $k=5$ nodes \cite{Przulj2004Graphlets}, then $k=6$ nodes \cite{Melckenbeeck1EtAl2016,melckenbeeck2017efficiently}, though BLANT is able to identify any graphlet up to $k=8$ nodes. Fig.~\ref{fig:graphlets} shows the complete list of $k$-graphlets for $k=3,4$. BLANT automates the process up to $k=8$, in which there are $n_k=12,346$ unique graphlets. Each $k$-graphlet is assigned a unique {\it graphlet ID} from $0$ to $n_k-1$ inclusive.

\begin{table}[htbp]
\begin{center}
\caption{Number of Graphlets and Unambiguous Graphlets}
\label{tab:num_graphlets_unambig}
\begin{tabular}{|r|r|r|}
\hline
\textbf{k} & \# \textbf{Graphlets}   & \# \textbf{Unambig.}  \\
\hline
2   & 1             & 0\\
3   & 2             & 0\\
4   & 6             & 0\\
5   & 21            & 0\\
6   & 112           & 8\\
7   & 853           & 144\\
8   & 11117         & 3552\\
\hline
\end{tabular}
% \captionsetup{width=1\textwidth}
%\caption{The total number of graphlets, and the number of unambiguous ones, for each value of $k$.}
% The suffixes K, M, G, T, P, and E represent exactly $2^{10}, 2^{20}, 2^{30}, 2^{40}, 2^{50}$ and $2^{60}$, respectively.}
\end{center}
\end{table}

An orbit is set of nodes that are topologically identical within a graphlet; {\it ie.,} they can be swapped without changing the graphlet. In Fig.~\ref{fig:graphlets}, nodes are shaded based on their orbit. For example, the first 4-node graphlet in Fig.~\ref{fig:graphlets} is the path of length 3, which has two orbits: the two middle nodes participate in one orbit, while the two end-nodes form another.

\begin{figure}
    \centering
    \includegraphics[width=0.4\textwidth]{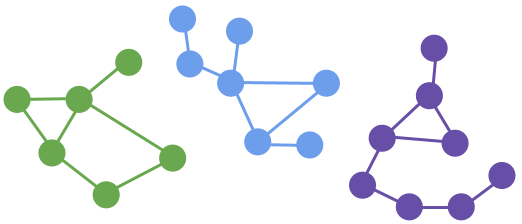}
    \caption{Examples of unambiguous graphlets on $k=6,7,8$ nodes. Each node in each graphlet exists in a unique orbit. No unambiguous graphlets exist for $k\leq 5$.}
    \label{fig:unambig_graphlets}
\end{figure}

A graphlet is defined as ``ambiguous'' if it contains at least two nodes of the same orbit. This is because there exist more than one way to create a topologically perfect alignment between two instances of an ``ambiguous'' graphlet, as any nodes of the same orbit may be swapped without affecting isomorphism. For example, when aligning two length-2 paths A-B-C and D-E-F, A and C may both be aligned with either D or F as the two endpoints of the path are in the same orbit. An ``unambiguous'' graphlet is simply a graphlet for which each node exists in a unique orbit. There is only one way to create a topologically perfect alignment between two instances of an ``unambiguous graphlet''. The smallest unambiguous graphlet has 6 nodes, as shown in Table \ref{tab:num_graphlets_unambig}. Some examples of unambiguous graphlets are shown in Fig. \ref{fig:unambig_graphlets}. To avoid the combinatorial explosion resulting from aligning ambiguous graphlets, from this point onwards we use only unambiguous ones—and note that there do not exist unambiguous graphlets with fewer than 6 nodes, meaning the ``traditional'' set of graphlets with up to only 5 nodes are insufficient.

\subsection{Alignments, and Types of Alignment Similarity}
\label{sec:alignments_and_sim_types}
\begin{figure}
    \centering
    \includegraphics[width=0.45\textwidth]{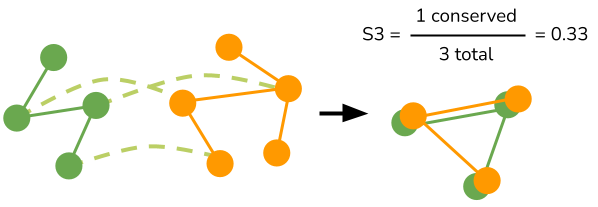}
    \caption{An figure explaining the symmetric substructure score ($S^3$). In the figure, an alignment of 3 node pairs is created between a graph of size 4 and a graph of size 5. The alignment contains 3 edges total. It contains 1 ``conserved edge'', or 1 edge which appears in both networks. Thus, the $S^3$ score is 0.33.}
    \label{fig:s3_example}
\end{figure}

We define an alignment as a 1-to-1 mapping from a set of nodes in one network to a set of nodes in another network. The number of nodes being mapped can range anywhere from 1 to the number of nodes in the smaller network. Fig. \ref{fig:s3_example} gives an example of an alignment of 3 node pairs between one network of 4 nodes and one network of 5 nodes.

The topological similarity of an alignment measures how close the aligned (sub)graphs are to being isomorphic. We use the measure the \textit{symmetric substructure score} ($S^3$, \cite{MAGNA}), which gives the fraction of the number of conserved edges in an alignment over the number of total edges in the alignment (cf. Fig. \ref{fig:s3_example}).

The functional similarity of an alignment relates to what the nodes and edges {\em are}, or what they {\em do}, in the real world. We make a simplifying assumption that each node has a single functional ``counterpart'' in the other network, which is mostly true in real world graphs. For example, in social networks, two nodes in different social networks are counterparts if they refer to the same person (1-to-1 except for duplicate accounts); in protein-protein interaction (PPI) networks, two proteins are considered identical if they are orthologs (related to common ancestors, mostly 1-to-1). When a node is aligned to its counterpart, we say the node pair is ``correctly aligned''. For functional similarity, we use the metric {\it node correctness} (NC), which is the fraction of node pairs in the alignment which are correctly aligned.

The reader should note the difference between ``information'' and ``similarity''. Our algorithm uses only topological information but is evaluated based on topological and functional similarity.

\section{Methods}
\subsection{Algorithm Overview}
\label{sec:algorithm_overview}
\begin{figure}
    \centering
    \includegraphics[width=0.47\textwidth]{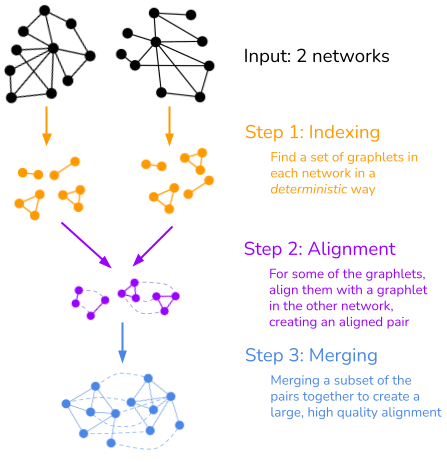}
    \caption{The three steps of BLANT. Note that our algorithm can be easily extended to more than two networks, but we use two for simplicity in this diagram. First, graphlets are deterministically extracted from each network individually. Then, pairs of identical graphlets from different networks are matched, creating a pool of small alignments. Finally, a subset of the pool is merged together into one final large alignment. }
    \label{fig:blant_pipeline}
\end{figure}

\noindent BLANT consists of three steps (cf. Fig.~\ref{fig:blant_pipeline}).

The first step is indexing, which takes in a single network as input and deterministically extracts a set of graphlets from the network. In the output, the graphlets are indexed by their graphlet ID. Each network only needs to be indexed once, and that index may then be used to attempt aligning the network with any other network which has been indexed.

The second step is alignment, which takes in two indexes as input and outputs a set of aligned pairs of identical graphlets which contain between 6-15 node pairs. The goal of our algorithm after this step is to have a set of graphlet pairs which have both high topological and functional similarity. Perfect topological similarity is enforced explicitly by aligning only identical graphlets. High functional similarity is achieved through the enforcement of perfect topological similarity, the use of determinism (cf. \S\ref{sec:introduction}), as well as the techniques described in \S\ref{sec:unambiguous_only}, \S\ref{sec:doubly_unique}, and \S\ref{sec:patch}.

The third and final step is merging, which takes in a set of aligned graphlets of 6-15 node pairs and outputs a single alignment of up to thousands of node pairs. While the aligned graphlets in the previous step were required to have identical topology, the large output alignment in this step does not need to be topologically perfect (we only enforce an $S^3$ score of 0.95, cf. \S\ref{sec:running_blant}). This step allows us to turn a large set of small, high quality alignments into a single large, high quality alignment. In this paper, we only aim to extract the largest and highest quality alignment we can from the set of aligned graphlets to demonstrate the effectiveness of our approach. We leave the problem of discovering multiple large, high quality alignments to future work.

\subsection{Step 1: Index Creation}
\label{sec:step_1_index_creation}
The basic idea of this algorithm is to recursively build graphlets by deterministically expanding from some root node. To ensure that every node has a chance to be in the index, this expansion is performed $n$ times, once with each node in the network serving as the root. The algorithm builds graphlets in a DFS-like fashion, creating a list of nodes it will search in each recursive step. It iterates through the list of nodes, adding each one individually to its working set of nodes and recursing from there before backtracking and adding the next node to the working set. In order to deterministically create this list, the algorithm selects neighboring nodes with the highest values according to a deterministic heuristic function. We control the exponential nature of recursive expansion by severely limiting the size of the node list at each recursive step.

\begin{algorithm}
\caption{Index Creation Algorithm}
\label{alg:index}
\begin{algorithmic}
    \Function{CreateIndex}{Graph $G$, $k$, $D$, $f$}
        \State sorted = $G$'s nodes sorted by $f$
        \State index = empty str -> [graphlet] map
        \For{$u$ in sorted}
            \State index += GetNodeEntries($G$, $k$, $D$, $f$, [$u$], index)
        \EndFor
        \State\Return{index}
    \EndFunction{}
    \State
    \Function{GetNodeEntries}{$G$, $k$, $D$, $f$, $V$, index}
        \If{$|V|=k$}
            \If{graphletID of $V$ is unambiguous}
                \State // Note: graphlets are stored as lists of nodes
                \State index[graphletID of $V$] = $V$
            \EndIf{}
        \Else{}
            \State $V_{exp}$ = GetExpandNeighbors($G$, $V$, $D$, $f$, index)
            \For{$u$ in $V_{exp}$}
                \State $V=V\cup \{u\}$
                \State GetNodeEntries($G$, $k$, $D$, $f$, $V$, index)
                \State $V=V-\{u\}$
            \EndFor
            \State\Return{entries}
        \EndIf{}
    \EndFunction
    \State
    \Function{GetExpandNeighbors}{$G$, $V$, $D$, $f$}
        \State neighs = \{all neighbors of all nodes in $V$\}
        \State uniqueValues = \{unique $f$ values in neighs\}
        \State expandValues = \{the $D$ largest values in uniqueValues\}
        \State $V_{exp}$ = \{subset of neighs with $f$ values in expandValues\}
        \State\Return{$V_{exp}$}
    \EndFunction{}
\end{algorithmic}
\end{algorithm}

\subsubsection{Selecting Neighbors to Expand to}
The index creation algorithm relies on a heuristic function, $f(v)$, to select which neighbors to expand to at each step. $f(v)$ assigns a topologically meaningful value to each node $v$ in the network in a deterministic way. We use $f(v)$=degree$(v)$ (with a slight modification) which we have found works well enough. Our slight modification deals with a practical problem of using degree: the expansion from each node approaches the hubs of the network relatively quickly, reducing the diversity of nodes in the output index. One solution is to have $f(v)$ ignore (meaning assign a value of 0 to) the $h$ highest degree neighbors at each expansion step, but this yields low performance because it is beneficial to \textit{eventually} expand to the hubs. Thus, we use a simple technique to resolve both these issues: we ignore the top $k - 1 - c$ highest degree neighbors at each expansion step, where $c$ is the number of nodes in our current set.

\subsubsection{Outputting Unambiguous Graphlets Only}
\label{sec:unambiguous_only}
The number of ways to align two graphlets of the same graphlet ID $G$ is:

\begin{equation}
\label{eq:num_align_ambiguous}
A(G)=\prod_o^{u(G)} {}_{f(G, o)} P_{f(G, o)}
\end{equation}

The number of unique orbits in $G$ is represented by $u(G)$, while $f(G, o)$ is the number of nodes in $G$ are in orbit $o$. Nodes of the same orbit may be permuted arbitrary in the alignment while maintaining isomorphism, hence the permutation operator. Further, for every permutation of nodes for one orbit, the nodes of any other orbit may be permuted in any way, hence why these permutations are multiplied together.

If any orbit has $f(G,o)>1$, then there are multiple ways to ``align’’ a pair of graphlets. As a simple example, a triangle can be aligned in 6 possible ways with another triangle: it can be rotated around a 3-cycle, then mirror-imaged and rotated 3 more times. As defined in \S\ref{sec:graphlets_orbits_ambiguity}, an unambiguous graphlet as one in which every node is its own orbit, eliminating all ambiguity in how to align it with another graphlet of the same ID.

Given our assumption about one-to-one node correctness (cf. \S\ref{sec:alignments_and_sim_types}), there is only a single correct way to align two graphlets. With a topology-only alignment algorithm, it is difficult to distinguish between different ways of aligning two graphlets of the same graphlet ID, as all of these ways yield topologically perfect alignments. Given that $A(G)$ grows combinatorially, the expected functional accuracy of aligning two graphlets of ID $G$ decreases rapidly as $f(G, o)$ increases. Fortunately, it is here that we are able to leverage BLANT's ability to perform ambiguity checking on graphlets of up to 8 nodes in $O(1)$ time \cite{hasan2017graphettes} in order to output only unambiguous graphlets, greatly improving functional accuracy.

\subsubsection{Time and Output Size Complexity}
\label{sec:index_time_complexity}
We utilize the standard output files from BLANT-sample \cite{blanttool,hayesblant}, based on \cite{hasan2017graphettes}, which contain information about graphlet IDs and orbits. We use these output files in order to perform the following operations in $O(1)$: converting a set of nodes to a graphlet ID and checking the ambiguity of a graphlet ID. Thus, we only perform $O(1)$ work in the base case.

In the recursive case, we visit all nodes with the top $D$ values of $f$, visiting at most $M$ nodes where $M\geq D$ ($M$ accounts for ties). Since the recursion depth is at most $k-1$, the time complexity for a search starting at one node is $O(M^{k-1})$. Since we call the recursive algorithm once per node, the overall time complexity of the algorithm is $O(nM^{k-1})$. This time complexity depends on the amount of ties in the heuristic function. For this paper, we focus on networks which exhibit a tail-heavy degree distributions (see Fig. \ref{fig:deg_distr}), which means that nodes with large degree have very few ties. Thus, $M$ will be approximately equal to $D$. Additionally, since $D$ and $k$ are both small, fixed constants in practice (cf. \S\ref{sec:parameter_selection}), the algorithm is fixed-parameter linear.

By definition, it is impossible to generate a file at a higher size complexity than the time complexity of the algorithm used to generate it, because the act of writing to a file is included in the time complexity analysis. Thus, the output size complexity is also bounded by $O(nM^{k-1})$, though it may be lower in practice due to duplicate in the output file.

\subsection{Step 2: Alignment}
The alignment algorithm takes in the indexes of any number of networks and mines a limited list of topologically identical $k$-graphlet alignments between them. We only output topologically identical alignments both because of the intuitive benefits discussed in \S\ref{sec:algorithm_overview} and because of the practical benefits: it allows us to simply use key equality to determine if two graphlets can be allowed, and it greatly simplifies the alignment process itself. Note that this step of the algorithm can be easily extended to take in more than two indexes, but we will use two for simplicity.

We utilize the standard output files from BLANT-sample \cite{blanttool,hayesblant}, based on \cite{hasan2017graphettes}, which contain information about the orbits of each graphlet ID. We use this information to align the nodes of two graphlets such that all aligned pairs are in the same orbit.

\label{sec:step_2_alignment}
\begin{algorithm}[t]
\caption{Alignment Algorithm}
\label{alg:align}
\begin{algorithmic}
    \Function{FindAlignedPairs}{File $F_1$, File $F_2$}
        \State // $F_1$ and $F_2$ are the output files of Alg. \ref{alg:index}
        \State $I$ = PatchIndex($F_1$)
        \State $J$ = PatchIndex($F_2$)
        \For{$k$ in $I$.keys $\cup$ $J$.keys}
            \If{$|I_k|=1$ \textbf{and} $|J_k|=1$}
                \State $S=S\cup \{(I_{k_0},J_{k_0})\}$
            \EndIf
        \EndFor
        \State\Return $S$
    \EndFunction{}
    \State
    \Function{PatchIndex}{$F$}
        \State $I$ = dictionary of empty sets
        \For{each line $l$ in $F$, where each line is a graphlet}
            \State $m$ = graphlet on line after $l$
            \If{$m$ and $l$ have any nodes in common}
                \State $p$ = graphlet from patching $m$ and $l$
                \State $k$ = ``patched graphlet'' ID of $p$
                \State $I_k=I_k\cup \{p\}$
            \EndIf{}
        \EndFor{}
        \State\Return $I$
    \EndFunction{}
\end{algorithmic}
\end{algorithm}

% WAYNEFIXME: I think we can remove the fact that BLAST automatically determines a good value of $k$. Getting high values of $k$ with graphs is so hard that we're never satisfied, so I think including this point would just distract the reader
%A: This one's important. We do NOT want ``as large as possible'', because the larger the graphlet, the more likely we'll hit an erroneous edge. There's a sweet spot between ``bigger is more unique'' and ``too big is likely to hit a difference between the networks that's not important''. BLAST makes the same trade-off with k-mers. Although we don't need to mention the error-vs-size thing, the ``theoretical'' statement about optimal value of k is important. Please put it back. Also, I think patching isn't about adding more volume by adding to the index; it's about creating higher k to get *unique* graphlets (but not too high, as above).
% We have found empirically that even $k=8$, the largest BLANT setting, does not have enough unique graphlets for our algorithm to perform well

\subsection{Doubly-Unique Keys}
\label{sec:doubly_unique}
%Index key specificity provides a highly effective way of pruning this space and increasing accuracy, usually at the expense of volume. %We have a constraint that the graphlets being aligned must be of the same graphlet ID, which gives us perfect edge-to-edge local alignments as output, and we leave the production of ``fuzzy'' alignments to future work.
In our index, the keys are graphlet IDs, which represent graphlet shapes. Say the graphlet shape $G$ appears $n_1$ times in the first index and $n_2$ times in the second index. Given our assumption about one-to-one node correctness (cf. \S\ref{sec:alignments_and_sim_types}), the maximum possible number of {\it correct} alignments is $\min(n_1,n_2)$. However, because it is difficult to prune these pairs when using only topology, we must output all $n_1n_2$ pairs. Assuming no node overlap in the alignments, our node correctness on these pairs is upper-bounded by: 

\begin{equation}
\label{eq:specificity}
\min(n_1,n_2)/(n_1n_2)
\end{equation}

Given this steep accuracy drop, we use the most stringent constraint of $n_1=n_2=1$, which we call ``doubly-unique keys''.

\subsection{Patching Graphlets}
\label{sec:patch}
%the performance of the algorithm depends on the number of keys indexed per network, as well an on the uniqueness of $k$-graphlets in a network. Just as with BLAST, larger values of $k$ increase the chance of having unique $k$-mers, but there is a tradeoff. Large $k$-mers and $k$-graphlets as sequencing errors or missing edges are more likely to appear. While BLAST automatically determines a good choice for this trade-off, the optimal value of $k$ for graphlets in our application is not yet known. However, we have found empirically that ``patching'' overlapping graphlets in the index into larger graphlets improves our results.

A significant challenge of using doubly-unique keys when keys represent graphlet shapes is that there are not that many different graphlet shapes for small values of $k$, especially when only considering unambiguous graphlets. As seen in Table \ref{tab:num_graphlets_unambig}, there are only 3552 different unambiguous graphlet shapes for the largest $k$ value BLANT is capable of, $k=8$, which is not enough keys in practice given the constraint of doubly-unique keys. In order to increase the number of unambiguous graphlets past what BLANT is natively capable of, we ``patch'' graphlets together \textit{after} building the index. We observe that BLANT outputs graphlets with a high degree of overlap in adjacent lines in the output index file, so we simply check every line against the line directly below it and patch the graphlets together if they contain at least one node in common.

\begin{figure}
    \centering
    \includegraphics[width=0.49\textwidth]{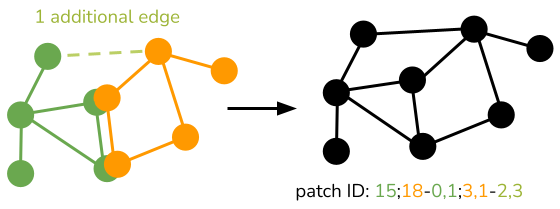}
    \caption{ Two 5-node graphlets with 2 nodes in common being patched together. The overlapping nodes are not being aligned; they are the exact same node in the exact same graph. The combined graphlet may also include additional edges that cross the two original graphlets. The patch ID is a string containing the graphlet ID of the first graphlet, the graphlet ID of the second graphlet, the orbits in each graphlet which overlap, and the additional edges. }
    \label{fig:patch_diagram}
\end{figure}

For further clarity, Fig.~\ref{fig:patch_diagram} shows an example of two graphlets being patched together. After patching, a new ``patch ID'' must be created that encodes the shape of the larger graphlet. Although this ID is not completely unique (all ID's refer to a single shape but multiple ID's may refer to the same shape), we have found that canonizing this ID using a tool like NAUTY \cite{NAUTY} decreases accuracy and volume. We hypothesize that this is because information about the two constituent graphlets positively informs the alignment because it retains how the deterministic algorithm expanded in that area of the graph. Additionally, while the patched graphlets may not be completely unambiguous, retaining the two constituent unambiguous graphlets always allows for an unambiguous alignment.

% Conveniently, our index creation algorithm provides the perfect way to find graphlets to patch together: because the algorithm performs DFS, graphlets with many nodes in common will be close to each other in the output file. Thus, we simply go through each line of the output file and compare the graphlet with the $P$ lines below it (cf. Fig.~\ref{fig:prox}). In addition to saving time, this method also increases accuracy as we are only patching graphlets which were discovered close to each other in the deterministic run. This result is demonstrated in \S\ref{sec:patching_parameters}.

% The second patching parameter, $C$, denotes the minimum number of nodes the graphlets must have in common to be patched. Smaller values of $C$ lead to larger graphlets, but the trade-off is that the patched graphlet will be less related to the graphlets in the index. The effects of $C$ are investigated in \S\ref{sec:patching_parameters}.

\subsection{Time Complexity}
\label{sec:patch_time_complexity}
As patching two graphlets takes $O(1)$ time, the step of creating the patched index takes $O(l)$, where $l$ is the number of lines in the larger index file. After creating patched indexes, the algorithm loops through the union of keys in both indexes and performs $O(1)$ work on each key, as it skips keys which are not doubly-unique. Since the total number of keys is bounded by $2l$, the total time complexity is $O(l)$. Since the size of the index file is bounded by $O(nM^{k-1})$ (cf. \S\ref{sec:index_time_complexity}), the time complexity of Algorithm \ref{alg:align} is also $O(nM^{k-1})$.

\subsection{Step 3: Merging}
\label{sec:step_3_merging}
The merging algorithms combines the smaller alignments created by the previous step into a larger local alignment. Before beginning the merging process, we filter out smaller alignments which are too globally similar, as described in \S\ref{sec:filtering_globally_similar_alignments}. Then, an iterative merging process begins. In each iteration, the merging algorithm randomly adds or removes an alignment $A_i$, with two constraints when adding. First, the overall topological similarity of the merged alignment $M$ must be above some threshold after adding $A_i$. Second, adding $A_i$ must retain a 1-to-1 matching between nodes in $M$. After some number of iterations, we terminate and output the largest alignment found thus far.

\begin{algorithm}[t]
\caption{Merging Algorithm}
\label{alg:merge}
\begin{algorithmic}
    \Function{Merge}{[[(Node, Node)]] $A$, float $m$, int $s$, float $t$}
        \State // $A$ is the list of alignments from Alg. \ref{alg:index}
        \State // $M$ is the merged alignment
        \State filter out alignments in $A$ with mean ODV sim. $>=m$
        \For{$s$ times}
            \State $i$ = random index of $A$
            \If{$A_i$ is in $M$}
                \State remove $A_i$ from $M$
            \Else
                \State $o2o=IsOne2One(M,A_i)$
                \State $s3'=IncS3(M,A_i)$
                \If {$o2o$ and $s3'\geq t$}
                    \State add $A_i$ to $M$
                \EndIf
            \EndIf
        \EndFor
        \State\Return largest $M$ found
    \EndFunction{}
    \State
\end{algorithmic}
\end{algorithm}

% TOREVIEW
\subsubsection{Filtering Out Globally Similar Alignments}
\label{sec:filtering_globally_similar_alignments}
As explained in \S\ref{sec:intro}, our goal is to discover non-trivially-sized regions within two networks which differ from the global alignment, yet are still topologically similar. A crucial technique which allows us to achieve this goal is filtering out alignments which are too globally similar. In order to predict global similarity using only topological information, we utilize the metric of orbit degree vector (ODV) similarity \cite{Milenkovic2008}. ODV similarity is defined between a pair of nodes, and is a value ranging from 0.0 to 1.0. As an alignment is a set of node pairs, we define the ODV similarity of an alignment as the mean ODV similarity of all of its node pairs. We filter out all input alignments which have a mean ODV similarity $>=m$.

There exists a fundamental tradeoff in the selection of $m$. The closer $m$ is to 1.0, the more similar the final merged alignment is to being a subset of the global alignment, which means it does not provide novel information. However, the further away $m$ is from 1.0, the more difficult it is to create large merged alignments with high topological similarity, because topological similarity and global similarity are correlated - the global alignment of any two networks is likely to contain far more topological similarity than a random alignment. We evaluate the optimal value of $m$ in \S\ref{sec:parameter_selection}.

\subsubsection{Time Complexity}
\label{sec:merging_time_complexity}
The merging algorithm involves a number of auxiliary data structures which reduce its time complexity. We will briefly describe them here. First, we keep a two sided map updated to represent $M$, allowing us to perform $IsOne2One()$ in $O(1)$ time. We also store a membership array which allows us to check ``$A_i$ is in $M$'' in $O(1)$ time. To calculate $IncS3()$, we store the current $S^3$ score of $M$ with both its numerator and denominator. We compare each pair in $A_i$ with each pair in $M$, incrementing the numerator and denominator according to the definition of $S^3$ (cf. Fig. \ref{fig:s3_example}). As the size of $A_i$ is constant, $IncS3()$ takes $O(p)$ time where $p$ is the number of pairs in $M$. The loop is repeated $k$ times, regardless of how many addition attempts failed. Thus, the total time complexity is $O(pk)$, where $p$ is the number of nodes in the final output alignment.

\section{Evaluation}
% the graph is how we show that no one is doing what we're doing
\label{sec:experiments}
In this section, we fix our algorithm's parameters and investigate its performance on a 21 large, realistic networks in the biological and social domains. We perform an in-depth comparison on our two testbeds against a baseline adapted from the non-topology-only local alignment setting. Finally, we analyze the time and space requirements of our index.

\subsection{Experimental Setup}
\subsubsection{Hardware and Environment}
All experiments are performed on a cluster of 96 identical machines (the ``circinus’’ cluster) in the Department of Computer Science at U.C. Irvine. Each host runs Linux CentOS, has 96GB of RAM and a 24-core Intel X5680 CPU running at 3.33GHz. Despite the large number of cores, the speed of a single core is comparable to that of a low-end laptop. Algorithm \ref{alg:index} is part of the larger BLANT package \href{https://github.com/waynebhayes/BLANT}{available on Github}. The index creation algorithm (Algorithm \ref{alg:index}) is part of BLANT and is written in C. The alignment algorithm (Algorithm \ref{alg:align}) and merging algorithm (Algorithm \ref{alg:merge}) are in Python. Our baseline, AlignMCL is also written in Python.

\subsubsection{Datasets}
\label{sec:datasets}
We evaluate our algorithm on two different testbeds of networks, listed in Table \ref{tab:networks}. First, we use the mammal PPI networks from the Integrated Interactions Database \cite{kotlyar2018iid}. The IID networks are, by far, the largest PPI networks available. Although they are partly synthetic, they are currently the best available approximation to ``real'' PPI networks. They (a) are nontrivial in size, and (b) share what is believed to be about the same amount of topological similarity as we expect in the (currently unknown) real networks.

Second, we use all but two of the temporal networks (incidentally all social networks) from the Stanford Network Analysis Project database, {\it SNAP} \cite{snapnets,kumar2018community,temporalNetMotifs,patternsDynamicsOnline,kumar2016edge,kumar2018rev2}. We ignore CommResistance because it is too small (fewer than 10 nodes per network). We ignore ActMOOC because it is bipartite, something we leave for future work.
It is important to note that our work does not utilize any special characteristics of temporal networks, a fact which we think demonstrates the generality of our approach. We have chosen to use temporal networks simply because they allow us to create perturbations of a network in a manner more realistic than simply removing edges at random.

\begin{table}[htbp]
\caption{Network Statistics}
\label{tab:networks}
\begin{center}
\begin{tabular}{|c|c|c|}
    \hline
    \textbf{Networks} & \textbf{\# Nodes} & \textbf{\# Edges} \\
    \hline
    All IIDs & 13K-18K & 256K-335K \\
    AskUbuntu 0\%-5\% & 20K & 49K\\
    BitcoinAlpha 0\%-5\% & 3.5K & 13K \\
    BitcoinOTC 0\%-5\% & 5.5K & 19K \\
    CollegeMsg 0\%-5\% & 1.0K & 5.5K \\
    EmailEUcore 0\%-5\% & 682 & 2.9K \\
    MathOverflow 0\%-5\% & 20K & 82K \\
    RedditHyperlinks 0\%-5\% & 20K & 60K \\
    StackOverflow 0\%-5\% & 20K & 191K \\
    SuperUser 0\%-5\% & 20K & 80K \\
    WikiTalk 0\%-5\% & 20K & 82K \\
    \hline
\end{tabular}
\end{center}
\end{table}

Each temporal network consists of a list of edges, each marked with a timestamp. To generate a network, we collect all edges between a start time and an end time, in what we call a ``temporal window''. We create four windows total, and each window contains the same number of edges\footnote{We determine this fixed number of edges by starting from the beginning of the temporal network and adding edges one by one until we either hit 20,000 nodes, 400,000 edges, an edge/node ratio of 20:1, or we run out of edges. We created windows based on nodes/edges instead of times because the distribution of edges over time was far from linear in many networks. This allows us to create networks at comparable sizes and densities as the IID networks.}. The first window begins with the earliest edge in the network. The second window begins by shifting the start time until we have ``lost'' 1\% of the edges, discounting duplicate edges which are ``regained''. The third and fourth windows use 3\% and 5\% shifting. The end times of each window are determined by collecting edges until we have hit the set amount.

\begin{figure}
    \centering
     % \begin{subfigure}
        \includegraphics[width=0.47\textwidth]{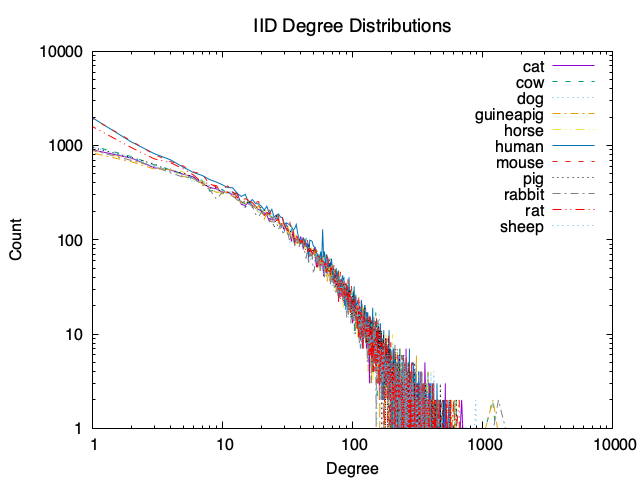}
    % \end{subfigure}
     % \begin{subfigure}
        \includegraphics[width=0.47\textwidth]{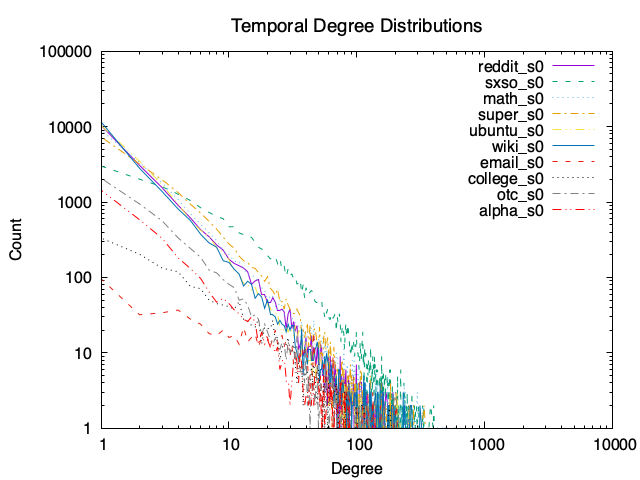}
    % \end{subfigure}
    \caption{Degree distributions of the IID networks (top) and temporal networks with 0\% shifting (middle). Only 0\% shifting is included because the degree distributions are essentially unchanged for 1-5\% shifting. The bottom diagram is a version of the middle diagram with count normalized. All three diagrams use a log scale for both axes.}
    \label{fig:deg_distr}
\end{figure}

All networks follow a tail-heavy degree distribution, as shown in the log-log graphs in Fig. \ref{fig:deg_distr}. This ensures that there will not be too many ties when choosing neighbors with the highest degree, keeping our algorithm's time complexity tractable (\S\ref{sec:index_time_complexity}).

In total, we use 11 IID networks and 10 temporal networks. We align every IID network against every other one, giving us 55 total IID network pairs. For each temporal network, we align the 0\% shifted with the 1\%, 3\%, and 5\% respectively, giving us 30 total temporal network pairs. In total, we consider 21 networks and 85 network pairs.

\subsubsection{Baselines}
To the best of our knowledge, BLANT is the first local aligner which was designed to use only topology. There exist many algorithms which first generate a global alignment using only topological information before mining that global alignment for local alignments. We discuss why we believe such algorithms are not ``true'' local alignment algorithms in \S\ref{sec:discussion} and \S\ref{sec:topo_only_global}. That said, we use one algorithm under this class, AlignMCL as a baseline.

AlignMCL was originally designed to use biological information, but it can be easily adapted to use only topological information as done in \cite{meng2016local}. The paper also adapted AlignNemo \cite{AlignNemo} to be topology-only, but we only use AlignMCL because the authors (the same for both algorithms) have stated that AlignMCL is the successor to AlignNemo. AlignMCL is a local alignment algorithm which uses the popular idea of combining two networks into a single alignment graph and then mining this alignment graph for higher quality local alignments. In order to create this initial alignment graph, AlignMCL utilizes the $n$ node pairs with the highest protein sequence similarity. Then, AlignMCL uses a Markov Clustering Algorithm to walk the alignment graph and discover clusters, which it outputs as local alignments.

In order to adapt AlignMCL to a topology-only setting, we used the approach used by \cite{meng2016local}: use the $n$ node pairs with the highest Orbit Degree Vector (ODV) similarity \cite{Milenkovic2008} (which we will call ``ODV pairs''), instead of protein sequence similarity. We initially got poor performance with this approach, which we discovered was because lots of nodes of degree 1 had identical ODVs and crowded out the other node pairs. We worked around this issue by simply ignoring degree 1 nodes when generating the the correctly aligned pairs.

\subsubsection{Metrics}
\label{sec:metrics}
We measure the alignments in terms of size (in nodes), functional accuracy, and topological accuracy. For functional accuracy, we use node correctness (NC, cf. \S\ref{sec:alignments_and_sim_types}), or the fraction of node pairs which are correctly aligned. For topological accuracy, we use symmetric substructure score ($S^3$, cf. \S\ref{sec:alignments_and_sim_types}), which measures the fraction of conserved edges in the alignment. Finally, we define a metric called ``alignment score'' which captures the overall ``quality'' of an alignment with the formula $score=n*NC^2*S3^2$. We square the accuracy values so that an alignment of 100 nodes and 50\% accuracy does not receive the same score as an alignment of 50 nodes and 100\% accuracy (``accuracy'' here conceptually includes both NC and $S^3$).

\subsection{Parameter Selection}
\label{sec:parameter_selection}
The parameters $k$ and $D$ in Algorithm \ref{alg:index} were straightforward to select. As mentioned in \S\ref{sec:patch}, the indexes of $k=6$ and $k=7$ contain far too many duplicate graphlets, as there are only 8 and 144 different unambiguous graphlets of size $k=6$ and $k=7$, respectively (cf. Table \ref{tab:num_graphlets_unambig}). Selecting a value of $D$ was also easy. $D=1$ is ruled out as it only outputs a single graphlet per node (assuming no ties in the heuristic function), since we only add one neighbor at each recursive step. $D=2$ already results in runtimes of 30-60 minutes for $k=8$ (cf. \S\ref{sec:runtime_and_storage}). As runtime increases exponentially with $D$ (cf. \S\ref{sec:index_time_complexity}), this essentially rules out $D>2$. Thus, we use $k=8$ and $D=2$.

\begin{figure}
    \centering
     % \begin{subfigure}
        \includegraphics[width=0.47\textwidth]{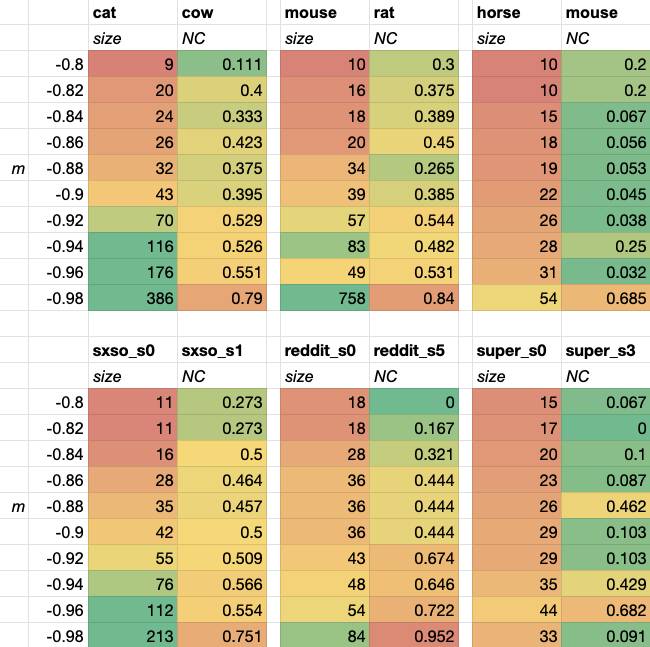}
    % \end{subfigure}
    \caption{How alignment size and node correctness (NC) vary based on $m$. The top half of the figure shows results from 3 IID network pairs, while the bottom half shows 3 temporal network pairs. Each group of two columns represents the results on a single network pair. We selected network pairs at varying levels of similarity, with similarity determined by our algorithm's performance on them. $s=5,000$ is used for all runs. Higher sizes are good (so higher size = more green) while higher NCs are bad (so higher NC = more red).}
    \label{fig:max_mean_odv}
\end{figure}

\begin{figure}
    \centering
     % \begin{subfigure}
        \includegraphics[width=0.47\textwidth]{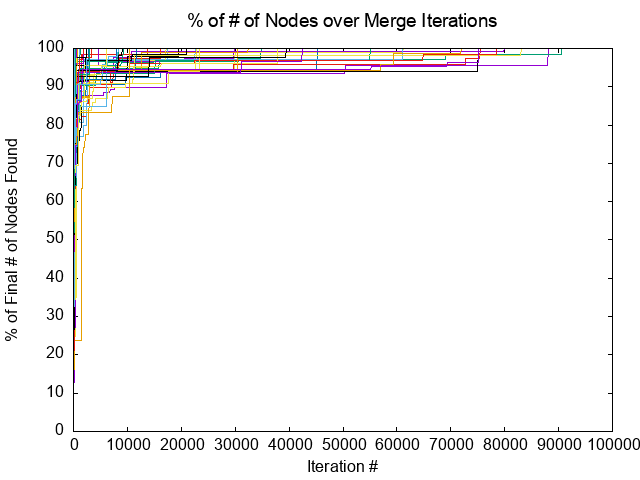}
    % \end{subfigure}
    \caption{The progression of the merging algorithm over iterations for all 85 network pairs. The value on the y-axis represents the number of nodes found in an intermediate point by the algorithm, as a percentage of the number of nodes found after 100,000 iterations. The x-axis represents iterations.}
    \label{fig:merging_iterations}
\end{figure}

The other parameters in our algorithm are $t$, $m$, and $s$ in Algorithm \ref{alg:merge}. As mentioned in \S\ref{sec:running_blant}, we choose $t=0.95$ to find alignments of high, but not perfect, topological similarity. To determine $m$, we selected 3 IID network pairs and 3 temporal network pairs and looked at both the alignment size and node correctness (NC) of the output alignment when running Algorithm \ref{alg:merge} with different values of $m$. As seen in Fig. \ref{fig:max_mean_odv}, the NC increases (bad) dramatically at around $m=0.96-0.98$ for all network pairs. Before this, the NC increases fairly slowly while the size increases (good) at a solid rate. Thus, we chose $m=0.95$ as a good tradeoff between size and NC. Finally, to determine $s$, we ran the merging algorithm with $s=100,000$ and observed how the largest alignment found so far increased across iterations, as shown in Fig. \ref{fig:merging_iterations}. As most network pairs converge quickly, we choose $s=20,000$. With this value of $s$, our Python implementation terminates in less than 1 minute on all network pairs.

% TODO: fix above paragraph to say $s$ and whatever the name of the max threshold parameter is

\subsection{Aggregate Performance Comparison}
\label{sec:aggregate_performance_comparison}
\subsubsection{Running and Postprocessing BLANT}
\label{sec:running_blant}
We run BLANT with the $S^3$ threshold of $t=0.95$, as the goal of local alignment is to discover a local region with significant, but not necessarily perfect, topological similarity. All other parameter choices are discussed in \S\ref{sec:parameter_selection}. After running BLANT, we take the largest connected alignment of BLANT's output alignment, where ``largest connected alignment'' is the largest alignment for which the induced subgraph of the nodes in one network are connected.

\subsubsection{Running and Postprocessing AlignMCL}
\label{sec:running_mcl}
We run AlignMCL with the top $n$ ODV pairs in each network, ignoring nodes of degree 1. AlignMCL has no other parameters. AlignMCL generates hundreds of alignments per network pair, and we process each of them the following way. First, as many alignments are not 1-to-1, we make them 1-to-1 by removing non 1-to-1 node pairs at random. Then, we take the largest connected alignment for each of their alignments.

\subsubsection{Comparison}
\label{sec:comparison}
\begin{figure}
    \centering
    % \begin{subfigure}
        \includegraphics[width=0.47\textwidth]{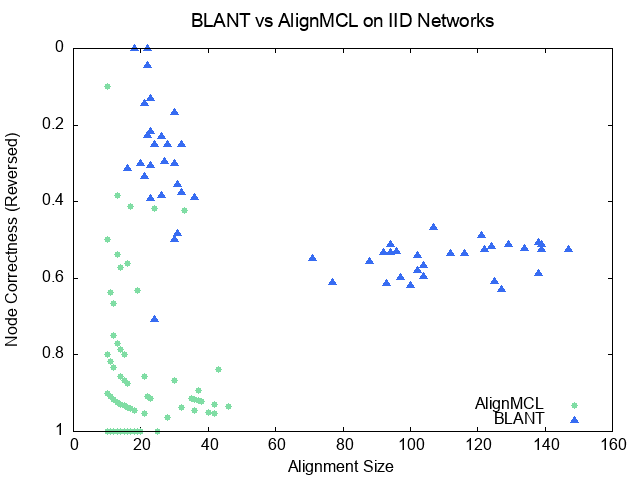}
    % \end{subfigure}
    % \begin{subfigure}
        \includegraphics[width=0.47\textwidth]{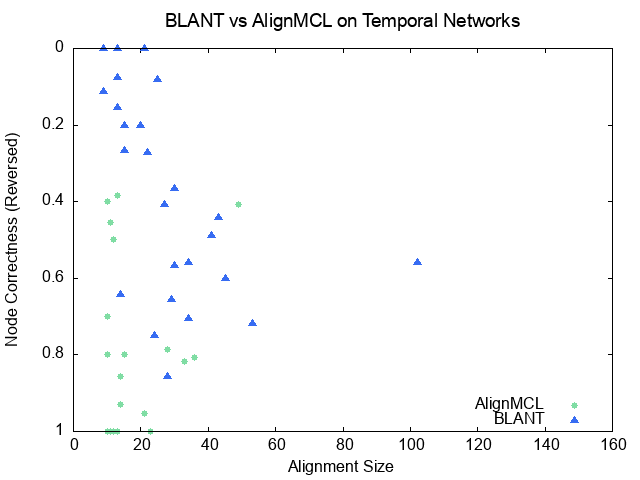}
    % \end{subfigure}
    \caption{Two plots comparing BLANT with AlignMCL in terms of size and node correctness on the IID network pairs and the temporal network pairs. All points shown have an $S^3$ score of $\ge 0.95$. The y-axis (node correctness) is reversed because our goal is to generate alignments of low node correctness.}
    \label{fig:blant_vs_mcl}
\end{figure}

In Fig. \ref{fig:blant_vs_mcl}, we compare the performance of BLANT and AlignMCL on the IID network pairs and the temporal network pairs, respectively.

All points in the plot represent alignments with $S^3 \ge 0.95$. BLANT generates a single high quality local alignment which is guaranteed to have $S^3 \ge 0.95$ because $t$ is set to 0.95 (cf. \S\ref{sec:parameter_selection}). On the other hand, AlignMCL generates hundreds of local alignments of varying quality per network pair. We filter out all AlignMCL alignments with $S^3<0.95$. Additionally, we filter out the significant number of alignments that have $<10$ nodes, because such alignments are trivial to find (any arbitrary aligned pair in Step 2 of our algorithm has 8-15 nodes).

\subsubsection{Discussion}
\label{sec:discussion}
On the IID networks, AlignMCL struggles to find large and topologically similar alignments with a low $NC$ (lower on the y-axis means higher $NC$). This makes sense, because their algorithm performs a step that resembles global alignment (computing global ODV pairs) before mining the global alignment for smaller local alignments. On the other hand, BLANT is able to discover large ($>100$ node) alignments with high topological similarity ($S^3>0.95$) that differ significantly from the underlying global alignment ($NC\approx 0.5$). Such local alignments are valuable because they provide different information than the global alignment.

On the temporal networks, BLANT does not perform as well as it does on the IID networks. However, the relative performance of BLANT compared to IID is the same; BLANT is generally able to find larger alignments that differ more from the underlying global alignment. Additionally, it is notable that AlignMCL is unable to find \textit{any} alignments with $S^3>0.95$ that consist of at least 10 nodes on most of the temporal network pairs. Specifically, it finds 0 alignments that fit these criteria on 19/30 of the temporal network pairs tested.

\subsection{Runtime and Storage}
\label{sec:runtime_and_storage}
\begin{figure}
    \centering
    %% \begin{subfigure}
        \includegraphics[width=0.47\textwidth]{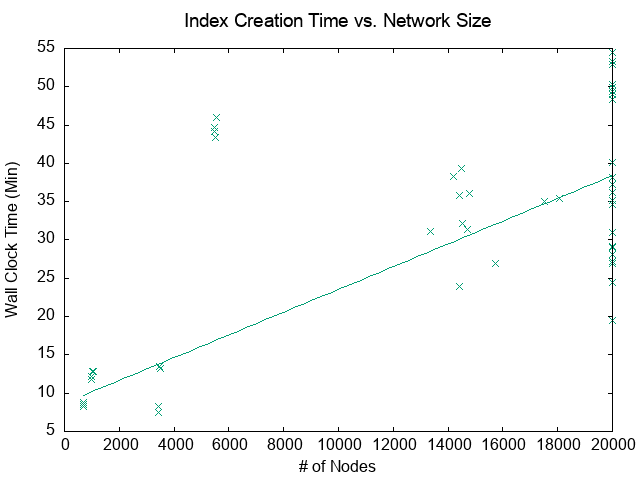}
    %% \end{subfigure}
    %% \begin{subfigure}
        \includegraphics[width=0.47\textwidth]{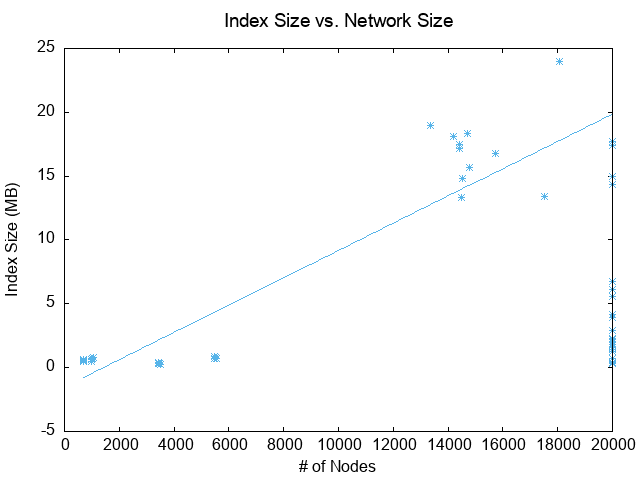}
    %% \end{subfigure}
    \caption{Plots showing index creation time vs. network size (top) and index size vs. network size (bottom). Index size is measured after removing duplicate lines. The trendlines are plotted with outliers removed. Both runtime and storage size grow approximately linearly.}
    \label{fig:time_and_space}
\end{figure}

With runtime, our goal is to be able to run the algorithm in a reasonable amount of time on a standard laptop. This measure is fuzzy as runtime is not a primary concern of ours, but we speculate that a researcher (the likely user of our algorithm) would feel that ``a few hours'' is a reasonable one-time cost to index a network. As shown in Fig. \ref{fig:time_and_space}, our algorithm's runtime grows linearly and takes less than an hour even for networks with 20000 nodes. Additionally, our index is fairly small, never exceeding 25MB even for the largest networks. We hypothesize that the significant variation in runtime is due to the different degree distributions of different networks, which results in different numbers of tied degrees at each expansion step (the effect of ties on time complexity is analyzed in \S\ref{sec:index_time_complexity}). Runtime and index size are fairly correlated, but differences between the two arise due to duplicate graphlets.

Other than indexing, we do not show figures. The alignment and merging steps take less than 5 minutes each despite being written in Python. Even though the time complexity of the alignment step is the same as that of the indexing step, the output indexes are very small in practice (cf. Fig. \ref{fig:time_and_space}), so the alignment steps runs very quickly.

We do not rigorously compare our runtime with that of AlignMCL, because the majority of AlignMCL's runtime is caused by a Python program we wrote which generates the list of seed pairs based on ODV similarity, which is in a different language than our indexing algorithm. Additionally, we do not perform a comparison because runtime is not a major concern of ours. To give some rough numbers, our Python program takes 5 hours to run for the largest graphs, and AlignMCL's own program takes 60 minutes for the largest graphs. This is comparable to the runtime of our index creation step, except our index creation step only needs to be run once per network while both parts of AlignMCL need to be once per network \textit{pair}.

\section{Related Work}
\label{sec:related_work}
\subsection{Topology-only Global Network Alignment}
\label{sec:topo_only_global}
Over the years, a number of global network alignment algorithms have been developed which rely solely on topological information. One broad approach is to start with some initial global alignment and randomly mutate it to improve some topological measure of similarity. MAGNA++ \cite{MAGNA++} does this with a genetic algorithm, while SANA \cite{MamanoHayesSANA} does this with simulated annealing. Other algorithms, like NATALIE \cite{NATALIE} and L-GRAAL \cite{LGRAAL}, model the global alignment problem as an integer linear program and use Lagrangian relaxation to produce a solution. Many machine learning approaches exist as well, such as CONE-Align \cite{Chen2020CONEAlignCN} and DANA \cite{dana2019}, which involve generating node embeddings and then aligning two embedding distributions in a lower dimensional space. In general, these global alignment algorithms process the entire graph as a whole while local alignment algorithms mostly utilize local information, showing the complementarity of these two approaches \cite{meng2016local}.

There exist algorithms, like GLAlign \cite{milano2018glalign} and the adapted version of AlignMCL \cite{meng2016local}, which combine these two approaches by first creating a global alignment and then breaking it up into multiple local alignments. However, the local alignments generated by these approaches can never differ significantly from the initial global alignment used to generate them, meaning the additional information they provide on top of the global alignment is limited (cf. \S\ref{sec:discussion}). Said another way, local alignment algorithms in this class do not complement global alignment algorithms well.

\subsection{Local Network Alignment with Side Information}
\label{sec:local_with_side}
``Side information'' refers to information outside of topology, such as node/edge attributes or pre-aligned seed nodes. Different domains tend to use different types of side information, demonstrating the difficulty of generalizing these algorithms. In aligning protein-protein interaction networks, local alignment algorithms may use information such as genomic sequence similarity \cite{mina2012alignmcl, AlignNemo} or COG function \cite{COOTES20071126}. In aligning knowledge graphs, algorithms use information such as entity name \cite{Ge2021LargeEAAE} or entity attributes (such as ``age'' or ``population'') \cite{RDFGraphAlignment}. In aligning social networks, algorithms most often apply percolation theory to grow alignments from a preexisting set of aligned seed nodes \cite{korula2014, kazemi2015}.

In some cases, local alignment algorithms which use side information can be converted into topology-only cases, as was done for AlignMCL \cite{mina2012alignmcl} and AlignNemo \cite{AlignNemo} in the comparison article \cite{meng2016local}. However, such conversions are not possible in many cases, as mentioned in \cite{meng2016local}. Additionally, these conversions often weaken some of the assumptions an algorithm is built on. For example, AlignMCL and AlignNemo rely on an auxiliary structure called an alignment graph which represents the merging of two graphs. The accuracy of this alignment graph degrades significantly when it is created with topological information instead of genomic sequence information, as seen in low performance of the topology-only version of these algorithms in \cite{meng2016local} and in \S\ref{sec:aggregate_performance_comparison}. This motivates the need for a local alignment algorithm which treats topology as a first-class citizen. Additionally, the topology-only global alignment algorithms have been demonstrated to produce complementary insights compared with global alignment algorithms which use side information \cite{GRAAL}. In future work, we hope to investigate whether this result applies to local alignment as well.

% MODULA: https://ieeexplore.ieee.org/stamp/stamp.jsp?tp=&arnumber=7359918 mine and merge, very similar to us

\subsection{Subgraph Querying and Indexing}
\label{sec:subgraph_querying}
The graphlet index we generate has many similarities, and differences, to the indexes produced by subgraph querying algorithms. Subgraph querying is the problem of finding all graphs in a database D which contain a query graph q as a subgraph \cite{lindex}. As this task involves solving subgraph isomorphism, an NP-hard problem, for each graph in D, a variety of heuristic algorithms exist. One prominent heuristic approach is ``filter-and-verification'' \cite{katsarou2015}, where a graphlet index is first created for all graphs in D, and this index is used quickly filter out graphs which definitely do not contain q. The remaining graphs, called the candidate set, are then verified to see if they contain q.

BLANT's graphlet index has many similarities with the indexes used for subgraph querying. \cite{katsarou2015} delineates four characteristics in the design space of indexes for subgraph querying that fully describe BLANT's index. Within this framework, BLANT's index stores graphlets as its features (characteristic \#1), mines its features non-exhaustively (characteristic \#2), uses the hash map data structure (characteristic \#3), and stores location information (characteristic \#4).

However, the main difference between the indexes of subgraph querying algorithms and BLANT's index is that the former is used for a non-existence check, while the latter is used for an existence check. Thus, indexes of subgraph querying algorithms need to be relatively exhaustive, while BLANT's index does not. Our core concept of determinism leverages this key insight and allows BLANT to overcome a significant issue with subgraph querying algorithms: the restrictively large index creation time and storage requirements. All algorithms studied in \cite{katsarou2015} have index creation times that grow exponentially or polynomially (as in, a polynomial > 1), and even the fastest evaluated algorithm, GRAPES \cite{Giugno2013GRAPESAS}, takes 3 hours just to index a graph of 2000 nodes. The storage requirements mostly grow exponentially or polynomially as well, and GRAPES comes with the tradeoff of requiring the most storage: 50GB for a graph of 2000 nodes. By contrast, BLANT's time and output size complexity are both fixed-parameter linear and BLANT takes 1 hour and uses 25MB to index graphs of 20000 nodes (cf. \S\ref{sec:runtime_and_storage}). By deterministically creating two indexes in the same way, we can obtain a similar slice of graphlets in two networks which contain similarity. Even though the number of distinct graphlets grows exponentially, the number of graphlets in the deterministic slice only needs to grow linearly.

\section{Conclusion}
To the best of our knowledge, we are the first to develop a local network alignment algorithm which relies solely on topological information. Our algorithm takes an entirely different approach from existing topology-only global alignment algorithms, as we only use local information instead of processing the graph as a whole. Our algorithm also takes a different approach from existing non-topology-only local alignment algorithms, as we use graphlets as our basic unit of local similarity instead of individual nodes. We utilize a key innovation, a deterministically generated graphlet index of the network, in order to prune the exponential search space of graphlets. This overcomes the restrictive runtimes of other graphlet indexes. Additionally, the use of determinism is crucial as it exploits the actual similarity---if present---among the set of networks in a way that exponentially reduces the search space. Then, we use numerous other techniques in order to query this index and expand the query results into large, high quality local alignments.

\section*{Acknowledgment}
We thank Brian Song and Ronit Barman for assisting with data collection and graph generation, and Arthur Jiejie Lafrance for implementing the heuristic function in the index creation algorithm.

\bibliographystyle{IEEEtran}
\bibliography{wayne-all}

\end{document}